\documentclass[preprint,11pt]{elsarticle}

\usepackage{amsmath,amssymb}

\usepackage{blkarray}
\usepackage{multirow}
\usepackage{easybmat,bm}

\usepackage{graphicx} 
\DeclareGraphicsExtensions{.bmp,.png,.pdf,.jpg}

\usepackage{array}
\usepackage{colortbl}
\definecolor{umbra}{rgb}{0.8,0.8,0.5}

\usepackage[table]{xcolor}

\begin{document}

\begin{frontmatter}




 \title{{\tiny \underline{Physica A 417 (2015) pp.176-184}}\\The Hurst exponents of \emph{Nitzschia} sp. diatom trajectories  observed by light microscopy}

\author[label1]{J. S. Murgu\'ia} \ead{ondeleto@uaslp.mx} 
\author[label2]{H. C. Rosu 
\corref{cor1}}\ead{hcr@ipicyt.edu.mx}
\author[label2]{A. Jimenez}\ead{andres.jimenez@ipicyt.edu.mx}
\author[label2]{B. Guti\'errez-Medina}\ead{bgutierrez@ipicyt.edu.mx}
\author[label3]{J.V. Garc\'{\i}a-Meza}\ead{jviridiana@gmail.com}

\address[label1]{Facultad de Ciencias, Universidad Aut\'onoma de San Luis Potos\'{\i},\\
  \'Alvaro Obreg\'on 64, 78000 San Luis Potos\'i, S.L.P., M\'exico}

\address[label2]{IPICYT, Instituto Potosino de Investigacion Cientifica y Tecnologica,\\
           Camino a la presa San Jos\'e 2055, Col. Lomas 4a Secci\'on, 78216 San Luis Potos\'{\i}, S.L.P., Mexico}

\address[label3]{Instituto de Metalurgia, Geomicrobiolog\'{\i}a, Universidad Aut\'onoma de San Luis Potos\'{\i}, \\
           \'Alvaro Obreg\'on 64, 78000  San Luis Potos\'{\i}, S.L.P., Mexico}

\cortext[cor1]{Corresponding author. Tel.: +52 (444)8342000, Ext. 718}

\begin{abstract}
We present the results of an experiment with light microscopy performed to capture the trajectories of live \emph{Nitzschia} sp. diatoms. The time series corresponding to the motility of this kind of cells along ninety-five circular-like trajectories have been obtained and analyzed with the scaling statistical method of detrended fluctuation analysis optimized via a wavelet transform. In this way, we determined the Hurst parameters, in two orthogonal directions, which characterize the nature of the motion of live diatoms in light microscopy experiments. We have found mean values of these directional Hurst parameters between $0.70$ and $0.63$ with overall standard errors below $0.15$. These numerical values give evidence that the motion of \emph{Nitzschia} sp. diatoms is of persistent type and suggest an active cell motility with a kind of memory associated with long-range correlations on the path of their trajectories. For the collected statistics, we also find that the values of the Hurst exponents depend on the number of abrupt turns that occur in the diatom trajectory and on the type of wavelet, although their mean values do not change much.\\

\noindent {\tiny Highlights:\\
$\bullet$ We determine Hurst parameters for digitally recorded 2D trajectories of a diatom species.\\
$\bullet$ The WT-DFA scaling method is used.\\
$\bullet$ We thus give statistical evidence for the persistent character of diatom motion.}

\end{abstract}


\end{frontmatter}







\section{Introduction}
\label{intro}

Ubiquitous in aquatic habitats, diatoms are photosynthetic unicellular microalgae that exist in both planktic and benthic lifestyles. The \emph{Nitzschia} species is a common genus of benthic diatoms that exhibits a paired mechanism of cell adhesion and gliding motility \cite{edgar1984,poulsen1999,wetherbee1998} (for a review see \citep{molino2008}). It is known that for such diatoms their adhesion and gliding is dependent on surface energies \citep{holland2004,li2010}, environmental chemical gradients \citep{wigglesworth1999} and growth phase \citep{deBrouwer2002,gupta2007}, as well as of a photosynthetic response  \citep{cartaxana2008,cohn1996,moroz1999,perkins2010b}.
Our purpose in the following is to discern independent features of their motility through a methodical examination of the dynamics of isolated cells in order to minimize the effect of a non-homogenous environment.

  \medskip

 At first sight and in the absence of external signals, the paths of migrating cells appear to be driven by a diffusive Brownian motion. However, in the experiment reported here, we focused on visualizing the individual \emph{Nitzschia} sp. trajectories with an automated tracking method developed in our laboratory. According to our observations, the diatom dynamics is characterized by alternating phases of directed gliding, changes of direction, and intermittency.
Individual cells and bodies which follow these dynamics have a prevalent non-Gaussian diffusion \citep{wang2012}.
Similar type of trajectories develop a multimodal search behavior that has been modeled and analyzed experimentally for different types of random motion \citep{li2008}, anomalous dynamics \citep{dieterich2008}, L\'evy distributions \citep{shlesinger1999}, intermittency \citep{benichou2006}, collective motion \cite{vicsek2012}, and also by direct observation of the diatom kinematics in constrained regions \citep{murase2011,umemura2013}. In this work, our viewpoint is that trajectories with more complex random walks, which are typical for cellular systems, can be approached by the underlying scaling laws of fractional dynamics \citep{shlesinger1999,ramanujan2006}.                    

\medskip

Our experimental observations are limited to spatiotemporal scales of hundreds of micrometers and seconds, which in general determine the common scales of motion of a single diatom. It is natural to think that the time series of the trajectories provide a means to understand the long-term movement of the diatom cells, which in fractal dynamics can be characterized by the Hurst exponent \citep{hurst1951} as a measure of spatiotemporal long-range correlations \citep{delignieres2013}. In this paper, we apply a fractal procedure based on the detrended fluctuation analysis (DFA) \citep{peng1994} to the time series of the trajectories of diatoms gliding freely. Based on the experimental results, we show that the migration of the \emph{Nitzschia} sp. diatoms is an active biological process dependent on the memory skills of these cells under the experimental conditions of our observations. The interpretation of the data as revealing persistent long-range correlations comes from the values of the Hurst parameter that we have obtained from the scaling-fluctuation coefficient $\alpha$ provided by the DFA.

\medskip

Before proceeding, we mention that in the recent paper \citep{murase2012}, a two-dimensional trajectory analysis of individual diatom cells of \emph{Navicula} sp. has been performed, and on the other hand, a correlation analysis similar to ours has been used in the area of completely sequenced genomes in \citep{knoch2009}.

\section{Brief description of the light microscopy experiment}
\label{s2}

\emph{Nitzschia} sp. cells have been isolated from biofilms in rock and soil surfaces that were submerged in a freshwater location near a thermoelectric power plant in San Luis Potos\'i, Mexico in the summer of 2012. Axenic subcultures were grown in Woods Hole medium at $25\,^{\circ}\mathrm{C}$ under a $14-10$ hrs light/dark cycles with cool white fluorescent light at the IM-UASLP Geomicrobiolog\'ia laboratory. This method for preparing Woods Hole culture media from freshwater algae and cyanoprokaryota has been adapted from the book chapter of Nichols \citep{nichols1973}.

\medskip

    Cells were inoculated and grown in either 24 well cell culture plates made of polystyrene (Costar; Corning Incorporated, catalog no. 3524, Corning, New York, NY)  or in $1$-L conical flasks with continuous oxygen supply. 
    Cultures grown in flasks were used throughout the experiments or otherwise noted and they contained \emph{Nitzschia} sp. diatoms that were able to perform and sustain both cell adhesion and diatom cell motility.
The diatoms have been slowly inoculated by capillarity inside a flow cell which immediately was wax-sealed and placed into the optical microscope mounting. The sealing enabled us to observe samples over several hours without internal currents that may perturb the assays. However, we noticed slightly higher motility rates when the samples were set for an incubation time of $90$ minutes after flowing diatoms into the flow cell.

\medskip

    The flow cell was placed on the specimen holder of an upright microscope to observe gliding \emph{Nitzschia} sp. diatoms which are about $30~\mu$m in size. Several diatoms can be recorded simultaneously in a single run, in our case, there were $2-17$ diatoms within each field of view of the microscope. After observing the sample for several seconds, diatoms gliding at approximately $5~\mu$m/s  were captured in videos at a rate of 26 fps. For analysis, video frames were digitized with VirtualDUB $1.9.11$ as uncompressed \texttt{avi} files, and each frame was exported as text images in ImageJ $1.46$ (National Institute of Health). For each sample, all data were collected within 2 h after preparation.

    \medskip

The object tracking was done with an in-house LabVIEW routine (National Instruments, Austin,TX). The digital video processing and tracking algorithm have been adapted by us to the experimental conditions of our lab and will be described elsewhere \cite{gutierrez2014}.
Briefly, the analysis of the two dimensional diatom trajectories starts by a tracking algorithm that determines the in-plane center-of-mass position of a single diatom and stores it for each video frame \cite{blumberg2005}. Raw positional data are subject to drift due to vibrations of the microscope and tracking errors. Thus, we applied a high pass filter and a moving average \cite{han2008} to correct for the drift, then we performed the scaling analysis to the drift-corrected time series. The trajectories were recorded for 5 minutes and only a few of them lasted less in the case of proximity to other diatoms which challenged the tracking routine.

 \medskip

    The trajectories of one hundred forty-one single diatoms were captured in video. Ninety-five (68\%) were circular-like and rather smooth with only a few derailments, while the other forty-four looked like failed attempts to achieve the circular-like motion and many of them still presented arcs of circles but in a more zigzag manner. For the WT-DFA analysis, we focused only on the ninety-five circular-like trajectories
    which looked as the most undisturbed, i.e., almost stationary for observations by light microscopy.

%

    \section{Time series of \emph{Nitzschia} sp. diatom trajectories}

    \subsection{Generalities}

  One frequent way of collecting experimental data is in the form of sequences of values at regularly spaced intervals in time.
  These sequences are called time series and provide useful information for further analysis and interpretation of the physical system that generated them.

  The stochastic nature of long-range dependence in a time series has been modeled by two interrelated processes: the \emph{fractional Brownian motion} (fBm) and the \emph{fractional Gaussian noise} (fGn).
  The fBm are Gaussian stochastic processes of zero mean, non-stationary and self-similar, which are indexed by the Hurst exponent $H \in (0,1)$, which in this work will be determined from the so-called scaling-fluctuation coefficient $\alpha$; see below.
The increments of fBm are stationary and are considered the basic example of an fGn process. Thus, fBm and fGn are invertible over integration and differentiation, respectively. An important feature of fBm is the presence of long-term correlations between the past increments and future increments. In a general sense, this indicates, on average, that the fluctuations on one time scale are statistically similar to the fluctuations
on other time scales.
  The Fourier analysis is a well established and suitable tool for analyzing stationary time series, whose statistical properties are not time dependent. In fact, the Fourier technique decomposes a signal into harmonic components because the basis functions are trigonometric functions. However, in recent times, it became evident that many time series are not stationary in the sense that their mean properties change in time.
  An alternative to the Fourier approach to deal with non-stationary time series is the wavelet transform (WT) \citep{mallat1998,lokenath2002,sifuzzaman2009}.
  The WT has been introduced and developed to study a large class of phenomena such as image processing, data compression, chaos, fractals, among
  others. The basis functions of the WT, the wavelets, have the key property of localization both in time (or space) and frequency, contrary to
  what happens with trigonometric functions. This wavelet property allows a more appropriate decomposition of non-stationary signals and their wavelet processing became a standard tool in the last two decades. However, to get the dynamics that produces a non-stationary signal it is crucial that in the corresponding time series a correct separation of the fluctuations from the average behavior, or trend, should be performed. Therefore, people had to implement novel statistical methods of detrending the data that should be combined with the wavelet analysis.
  At the present time, there are several such methods and techniques that have been developed to analyze non-stationary time series that display singular behavior or multiple scaling behavior of multifractal type.

  \subsection{DFA Algorithm}

  Originally proposed in 1994 by Peng \textit{et al} \citep{peng1994}, the DFA method can be applied to both fBm and fGn.
  Its advantage is that it outperforms over other more conventional techniques in quantifying the correlation properties of intrinsic self-similarity embedded in a seemingly nonstationary time series \citep{peng1994,K02,Eke}. In addition, it also avoids the spurious detection of apparent self-similarity, which may be an artifact of extrinsic trends.
  With the aim to have a very acceptable algorithm for the DFA method with much less computational cost and better accuracy,
  in 2005 Manimaran \textit{et al} \citep{mani} proposed to combine the orthogonal discrete wavelet transform with the DFA procedure.
  Based on this, Murgu\'ia \textit{et al} \citep{murguia-AMF-AC} implemented and used this approach to investigate the multifractal behavior of the time series of the row-sum signals of certain cellular automata rules \citep{murguia-AMF-ACs}.
  This method is known under the acronym WT-DFA and we use it for the diatom data because of the experience gained with its usage in our previous papers. The WT-DFA basically exploits the fact that the low-pass version resembles the original data in an ``averaged''
  manner in different resolutions. Instead of a polynomial fit, we consider the different
  versions of the low-pass coefficients to calculate the ``local'' trend. The numerical analysis of a time series $x(t_k)$, where $t_k=k \Delta t$ and $k=1,2,\ldots,~N$, within the framework of the WT-DFA procedure is performed in the following steps:

  \begin{enumerate}

    \item  Determine the profile $Y(k)=\sum_{i=1}^{k} (x(t_i)-\langle x\rangle )$ of the time series, which is the cumulative sum of
    	the series from which the series mean value is subtracted.

    \item Compute the fast wavelet transform (FWT), i.e., the multilevel wavelet decomposition of the profile.
          For each level $m$, we get the fluctuations of the $Y(k)$ by subtracting the ``local'' trend of the $Y$ data, i.e., $ \Delta Y(k;m) = Y(k) - \tilde{Y}(k;m)$, where $\tilde{Y}(k;m)$ is the reconstructed profile after removal of successive details coefficients at each level $m$. These fluctuations at level $m$ are subdivided into windows, i.e., into $M_s={\rm int}(N/s)$ non-overlapping segments of length $s$. This division is performed starting from both the beginning and the end of the fluctuations series (i.e., there are $2M_s$ segments). Next, one calculates the local variances associated to each window $\nu$
         \begin{align}\label{eq-Fs1}
           F^2(\nu,s;m) & ={\rm var}\Delta Y((\nu-1)s+j;m), \\
           j & =1,..., s~, \quad \nu=1,..., 2M_s~, \quad M_s={\rm int}(N/s)~. \notag
         \end{align}

    \item  Calculate the fluctuation function defined as
          \begin{equation}\label{eq-Fqs}
            F_2(s;m) = \left( \frac{1}{2M_s} \sum_{\nu=1}^{2M_s} |F^2(\nu,s;m)| \right)^{1/2}.
          \end{equation}

	\item Repeat the above procedure for different (a broad range of) segment lengths $s$.

  \end{enumerate}

    If the fluctuation function $F_2(s;m)$ displays a power law scaling
       \begin{equation}\label{eq-FqsPLaw}
         F_2(s;m) \sim s^{\alpha},
       \end{equation}
 then the analyzed time series have a fractal scaling behavior, with scaling-fluctuation exponent $\alpha$.
   This exponent can be found as the slope of the line expressing \eqref{eq-FqsPLaw} in the $\log F_2$ versus $\log s$ plot, and it
   is a measure for the degree of correlation in the time series. If $\alpha  = 0.5$ there is no correlation
   and the signal is uncorrelated (white noise). On the other hand, if $0 < \alpha < 0.5$ the signal presents an anticorrelated behavior
   (alternation between small and large values), and the time series is said to be anti-persistent;
   if $0.5 < \alpha < 1$, thus the correlations in the time series are persistent, where large values in
   the series of data are more probably to appear after large values, and vice versa. The values
   $\alpha = 1$ and $\alpha = 1.5$ correspond to $1/f$-noise and Brownian motion, respectively.
   This exponent can be considered as a generalization of the Hurst exponent. For stationary time series,
   $\alpha$ is identical to the Hurst exponent $H$, whereas for non-stationary time series $\alpha=H+1$, \citep{delignieres2006,K02,Eke}.
   This scaling-fluctuation exponent is also related to other exponents, such as the scaling exponent $\beta$ of the Fourier spectral density of the signal (the power spectrum) by $\alpha = (1+\beta)/2$.	

   For the fBm, the relationship is $\beta = 2H + 1$, where $1 < \beta < 3$, since $H$ lies between
   0 and 1 and classical Brownian motion is a special case corresponding to $\beta= 2$.
   On the other hand, for the fGn, $\beta = 2\alpha - 1$, and $\beta = 2H - 1$, with $-1 < \beta < 1$, for example the
   classical white stationary Gaussian noise is a special case with $\beta = 0$ \citep{delignieres2006,buldyrev1995}.


 \subsection{Hurst exponents of time series of \emph{Nitzschia} sp. diatom trajectories recorded for 5 minutes}

\begin{figure}[!ht]
  \centering
  \includegraphics[width=0.75\textwidth]{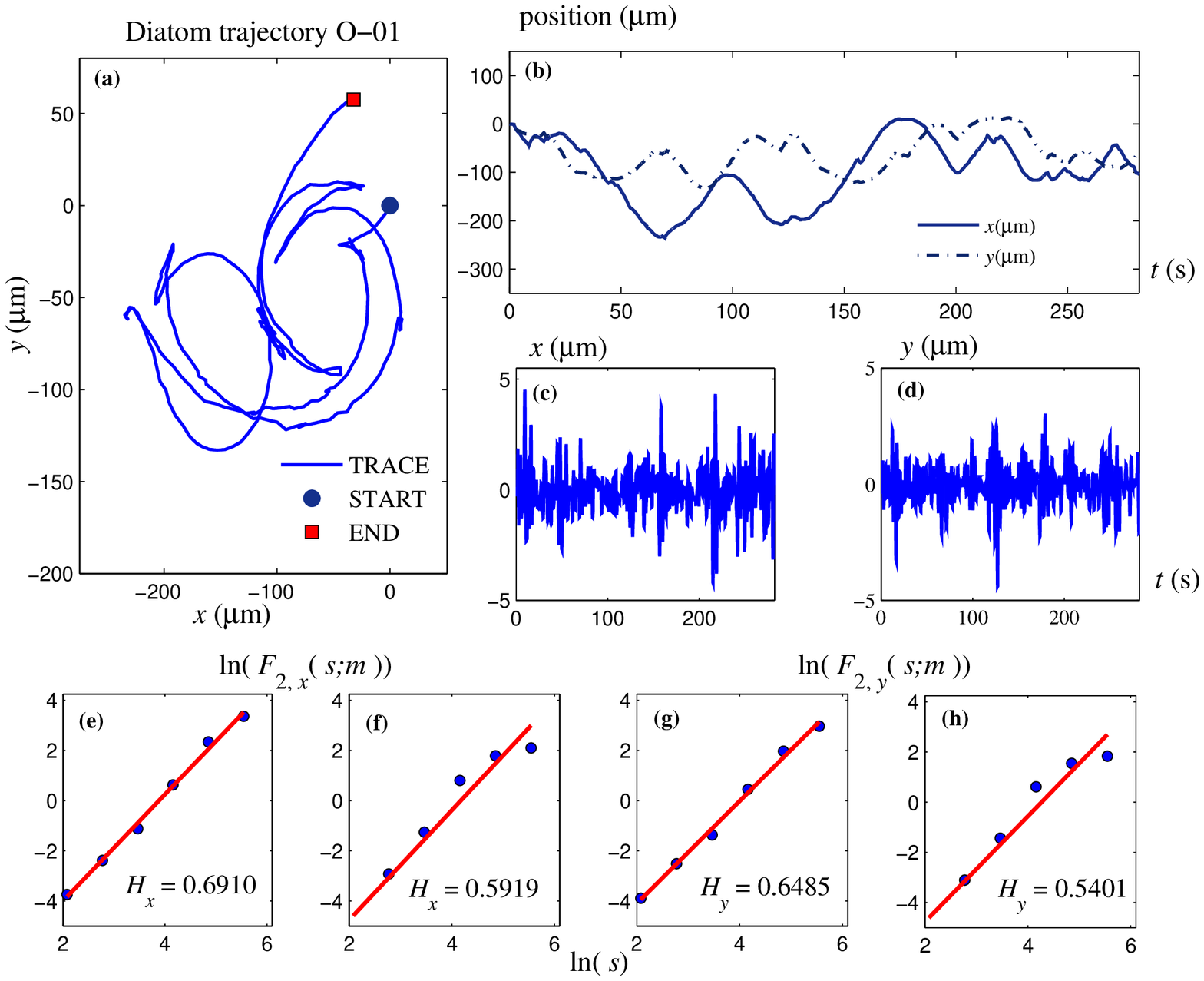}
  \caption{Diatom trajectory (a), the measured $x$ and $y$ time series (b), the same time series after applying the SVD method of eliminating the periodic trend has been applied (c,d), and bilogarithmic plots of $F_{2,x}$ (e,f) and $F_{2,y}$ (g,h) for db-3 and db-4 wavelets, respectively.
  	The Hurst exponents of this trajectory are $H_x = 0.6910$ and $H_y = 0.6485$ in the case of db-3 wavelet and $H_x = 0.5919$ and $H_y = 0.5401$ in the case of db-4 wavelets.}
  \label{fig-mf2}
 \end{figure}

\begin{figure}[!ht]
  \centering
  \includegraphics[width=0.75\textwidth]{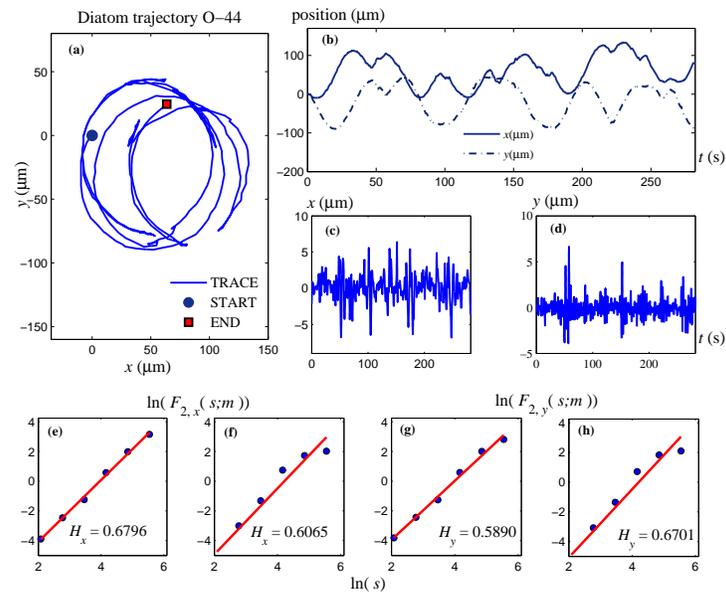}
  \caption{Same as in the previous caption but for another trajectory.
  	The Hurst exponents are $H_x = 0.6796$ and $H_y = 0.5890$ for db-3 wavelets and $H_x = 0.6065$ and $H_y = 0.6701$ for db-4 wavelets.}
  \label{fig-mf5}
 \end{figure}

    We now report our WT-DFA results for the recorded circular-like \emph{Nitzschia} sp. trajectories which looked as the most stable and therefore as the most natural for these diatoms under the experimental conditions of light microscopy.
    We used the db-3 and db-4 wavelet functions belonging to the Daubechies orthogonal family because this family has a number of desirable properties, such as orthogonality, approximation quality, and numerical stability \cite{lokenath2002,mallat1998,mani}.
    In addition, the WT-DFA algorithm with the Daubechies wavelet family is memory efficient and is reversible, whereas other wavelet
    bases have a slightly higher computational overhead and are conceptually more complex.
    The results of this scaling method for the correlation properties of the drift-corrected time series are presented in two selected
    figures, Figs.~\ref{fig-mf2} and \ref{fig-mf5}, not to burden the paper with too many plots. Since we wish to accurately estimate the Hurst parameters in two orthogonal directions, which for circular-like trajectories display (quasi)periodic patterns, we eliminate this periodic trend by means of the singular value decomposition (SVD) method, which is a common technique used in the literature for this purpose \cite{N05}. We implemented the SVD algorithm based on \cite{N05} using a fixed value of the dimensionality parameter, while the number of frequency components was chosen the same as the order of the employed wavelet. In this way, we achieved a satisfactory numerical stability of the Hurst mean values and on the other hand it is natural to think that for quasi-circular trajectories more harmonics get involved. The SVD filtering has been applied prior to the WT-DFA estimation because it is long known that the DFA algorithms cannot deal with periodic trends \cite{hu01}.


    \medskip

To investigate any dependence of the $H$ exponents with the number of abrupt turns in a diatom trajectory and with the type of wavelets, a first statistical evaluation of normality is provided in Table \ref{tab:diatomA1} which was obtained with the Shapiro-Wilk normality test that we performed using the package SPSS v.20 (IBM, Armonk, NY).

\begin{table}[h!]
\caption[xxx]{The mean values $\mu$ and the variances $\sigma$ for the distributions of time lags between successive abrupt turns and the Hurst exponents of the global statistics in both directions for two different Daubechies wavelets. The output $p$-values of the Shapiro-Wilk normality test for these distributions are displayed in the last column.\\}
		\label{tab:diatomA1}
	\centering
	\begin{tabular}{ccccc}
	\hline
{Distribution} & {N} & {$\mu$} & {$\sigma$} & {\emph{p}}\\
\hline
{$\Delta t$}	& {488}	& {26.094 s}	& { 0.038 s}	& {0.001}\\
{$H$ (db-3)}	& {190}	& {0.718}	& {0.123}	& {0.001}\\
{$H$ (db-4)}	& {190}	& {0.659}	& {0.145}	& {0.001}\\
\hline
\end{tabular}
\end{table}

Each trajectory displays none or several abrupt turns whenever the diatom changes its direction of motion with corresponding elapsed times between the turns $\Delta t_0$, $\Delta t_1$, $\Delta t_2$, $\dots$, $\Delta t_n$. To determine each $\Delta t_i$, the time series and the video recorded trajectory were used in synchrony for all trajectories.

The \emph{p}-values obtained by the Shapiro-Wilk normality test indicate that the $\Delta t$ distribution and the global $H$ distributions
are not normal distributions. For Hurst statistics of individual directions similar results have been obtained. This non-normality feature may be due to some bias effects or to the fact that all the statistical quantities are estimated from each random single-particle trajectory for which the estimators are not necessarily Gaussian even in the limit of very large number of collected data. This is well known in the literature of single-particle tracking measurements for the similar case of the estimators of the diffusion coefficient of Brownian or fractional Brownian motion \cite{greb1, greb2}. Future advanced experiments may shed more light on the origin of this non-normality.

\medskip

\begin{figure}[!ht]
  \centering
  \includegraphics[width=0.75\textwidth]{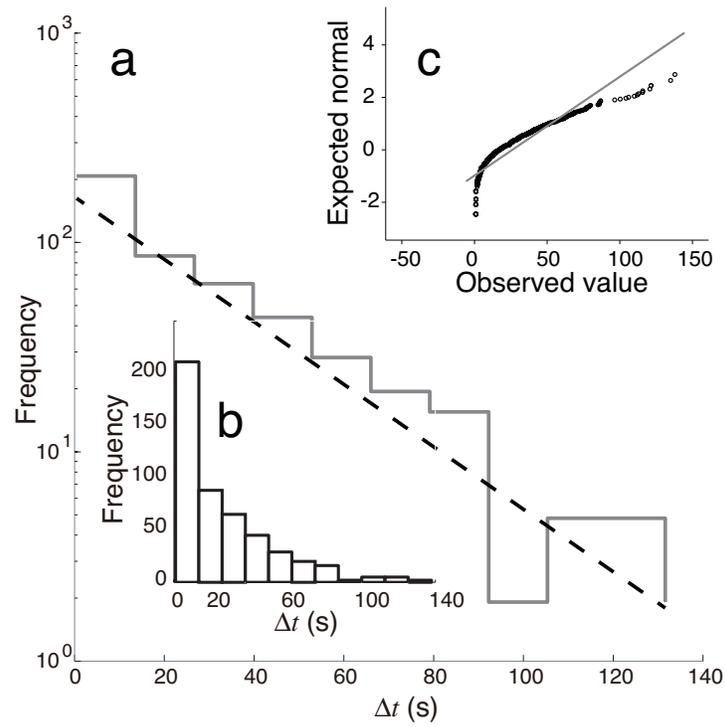}   
  \caption{(a) Histogram of time intervals $\Delta t$ between abrupt turns occurring in each trajectory fitted by an exponential distribution shown in solid and dashed lines, respectively, with characteristic time of 19.626 s. (b) Same histogram but in linear scale. (c) The normal Q-Q plot of $\Delta t$.}
  \label{fig-mfA1}
 \end{figure}

Figure \ref{fig-mfA1} shows the distribution of $\Delta t$ and its normal Q-Q plot. The normal Q-Q plot compares how the observed values of $\Delta t$ distribute along the expected normal and confirms, in this case, that the time lags between successive abrupt turns deviate from a normal distribution. This kind of distribution is characteristic of biological Poisson processes suggesting that the time condition that triggers an abrupt turn in \emph{Nitzschia} sp. is a stochastic memoryless process. The time constant of the process in the case of this diatom is 19.626 s.

\medskip

\begin{figure}[!ht]
  \centering
  \includegraphics[width=0.75\textwidth]{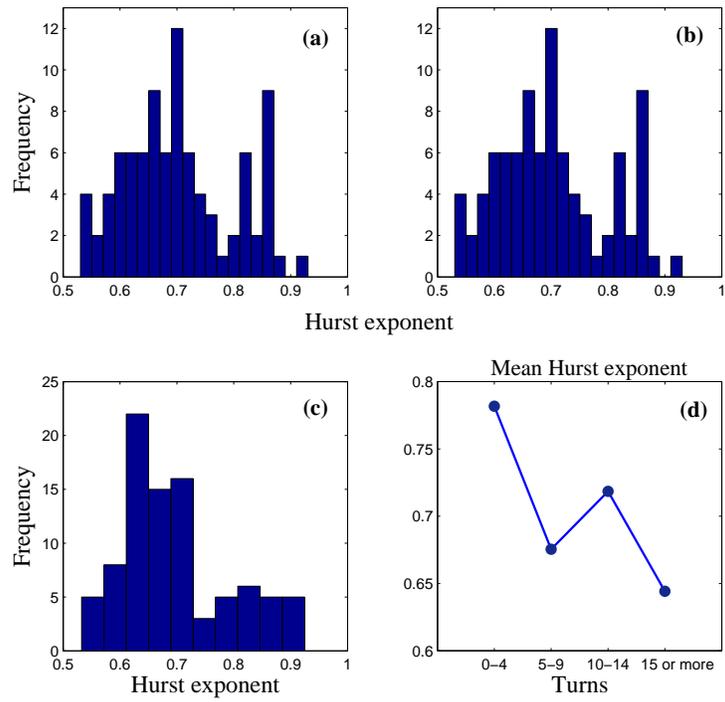}
  \caption{The distribution of the Hurst exponents of 90 diatom trajectories obtained with the db-3 wavelets in
  (a) $x$-direction, (b) $y$-direction, and (c) both directions together. (d) The dependence of the global Hurst exponents with the number of turns.}
  \label{fig-mf6}
 \end{figure}

\begin{figure}[!ht]
  \centering
  \includegraphics[width=0.75\textwidth]{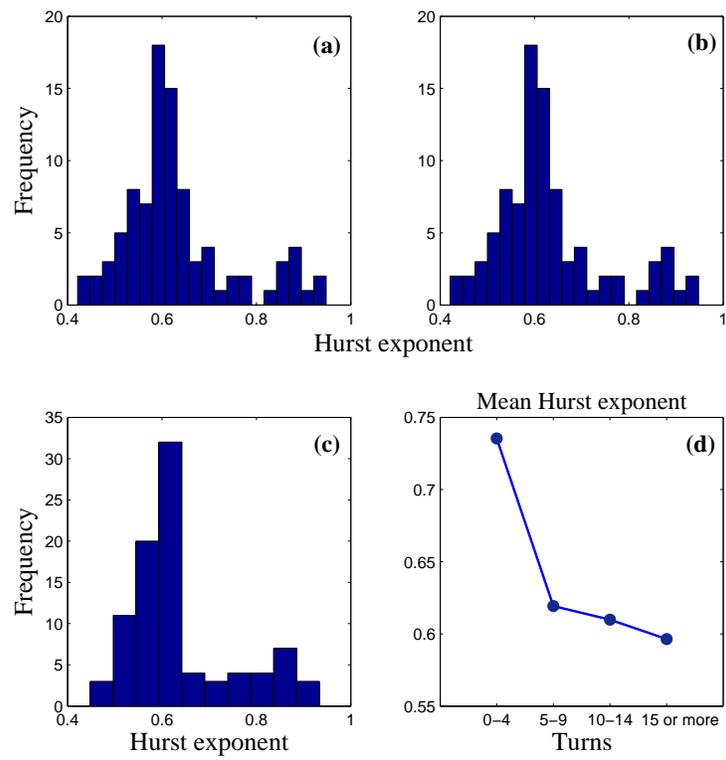}
  \caption{The same as in the previous figure but for 91 trajectories in the case of db-4 wavelets.}
  \label{fig-mf7}
 \end{figure}

  The distributions of the Hurst exponents in the two orthogonal directions corresponding to the circular-like trajectories after eliminating several outliers are presented in the histograms of Figs. \ref{fig-mf6} and \ref{fig-mf7} for the db-3 and db-4 wavelets, respectively.
  The mean values of these distributions are $\overline{H}_x=0.7031\pm 0.0964$ and $\overline{H}_y=0.6985\pm 0.1063$ for the db-3 wavelets and
  $\overline{H}_x=0.6314\pm 0.1153$ and $\overline{H}_y=0.6521\pm 0.1244$ in the case of db-4 wavelets. We also performed calculations with db-5 and db-6 wavelets and noticed stable results, such as $\overline{H}_x=0.6973\pm 0.0998$ and $\overline{H}_y=0.7007\pm 0.1071$ in the first case and
  $\overline{H}_x=0.6884\pm 0.1116$ and $\overline{H}_y=0.6952\pm 0.1154$ in the second case. However, all these stable mean values have been obtained by increasing the number $p$ of frequency components in the SVD method from 3 to 4, 5 and 6 for db-3, db-4, db-5, and db-6 wavelets, respectively, while keeping the SVD embedding dimension parameter fixed ($d=110$).

Moreover, four levels of abrupt turns, namely 0-4, 5-9, 10-14, and 15-or-more, are defined to group trajectories by their number of abrupt turns, i.e., the level of turns 0-4 is the set of trajectories with a number of abrupt turns ranging from zero to four and so forth, and
we applied an ANOVA two-way analysis of variance for two factors: level of turns and wavelet choice. The results of this analysis are summarized in Table \ref{tab:diatomA2}.

\begin{table}[h!]
\caption[xxx]{The two-way ANOVA calculated values. \emph{f}$_{critical}$ corresponds to a level of significance of {5}\%.\\}
		\label{tab:diatomA2}
	\centering
	\begin{tabular}{cccccc}
	\hline
& {SS} & {\emph{df}} & {\emph{f}} & {\emph{p}} & {\emph{f}$_{critical}$}\\
\hline
{Level of turns} & {0.0182} & {3} & {14.2546} & {0.0279} & {9.2766}\\
{Wavelet choice} & {0.0084} & {1} & {19.7778} & {0.0212} & {10.1280}\\
{Error} & {0.0013} & {3} & {} & {} & {}\\
{Total} & {0.0279} & {7} & {} & {} & {}\\
\hline
\end{tabular}
\end{table}

The resulting set of $f_{3,3} = 14.2546$ and $p = 0.0279$ is not statistically significant. Therefore, it fails to reject the chosen null hypothesis that $H$ does not depend on the level of turns. In addition, $f_{1,3} = 19.7778$ and  $p = 0.0212$ are not statistically significant either, and therefore, does not reject the null hypothesis that the global $H$ distribution does not depend on the choice of wavelet. For individual $H$ distributions similar results are obtained.

To quantify the difference between the wavelet of choice used in this work a Wilcoxon-Mann-Whitney U test was performed, which is known to be more efficient than the $t$-test \cite{fay10}. We found a U value of 11909.000 and a $z$-score of $-5.736$ ($p=0.001$), which is statistically significant at a level of significance of $5$\%, meaning that one wavelet ranks higher than the other. For our data, the assigned mean ranks for the total statistics in both directions are of $222.820$ and $158.180$ for db-3 and db-4, respectively. Similar results are obtained for the individual $x$ and $y$ statistics.

 \section{Conclusions}

 We have applied the WT-DFA scaling analysis to the diatom motion using the time series of the \emph{Nitzschia} sp. diatom trajectories digitally recorded for 5 minutes in observations by light microscopy. From this analysis, we have determined the values of the Hurst exponents corresponding to 95 time series of circular-like trajectories of this diatom species. After deleting a few outliers, the mean values of the Hurst exponents in the two orthogonal directions are between 0.70 and 0.63 with standard errors below 0.15 which point to
 a persistent type of motion. These values indicate some sort of memory-type effects but more experiments of this type with better statistics are needed to quantify and calibrate the motion of different diatom species in terms of scaling parameters.

 The values of the Hurst exponents depend on the number of abrupt turns observed in a \emph{Nitzschia} sp. trajectory with a characteristic time of 19.626 s and also on the type of wavelet. To settle the issue of these dependencies a bigger amount of experimental data is needed. We also confirmed with the Wilcoxon-Mann-Whitney U test that the choice of order of the Daubechies orthogonal wavelets, db-3 and db-4, influences the exponent values, getting higher ranks in the first case. The results obtained with the db-3 wavelets can be judged as more reliable because these wavelets have a smaller support size which increases their ability to display the non-smooth characteristics of the diatom trajectories and at the same time they have an acceptable number of vanishing moments required to improve the multiresolution analysis for the smoother part of the trajectory signals.\\

\bigskip
\bigskip

\noindent Acknowledgements\\

\noindent A. Jimenez thanks CONACyT for a graduate fellowship and IPICyT for financial support. We also wish to thank the referees for very useful remarks. J.S. Murgu\'{\i}a wants to dedicate this paper to the memory of his former colleague and friend at UCSD, Dr. Alexander Vergara, who passed away in March 2014.\\





\end{document}